\documentstyle[12pt,epsf]{article}

\begin{document}
\begin{center}
{\large {\bf Resonance states in the $^{12}$C+$^{12}$C modified Morse potential}} \\
\end{center}
\vskip 0.5 true cm \centerline {L. Satpathy} \centerline {\it
Institute~ of~ Physics,~ Bhubaneswar - 751 005,~India}

\vskip 1.2 in

\centerline {\Large {Abstract}}

The resonances in $^{12}$C+$^{12}$C system described earlier using
long range Morse potential determined from resonance data itself,
are reexamined in the light of the recent development of the two
new methods for identification of resonances in the scattering
theory, namely Imaginary Test Potential and Imaginary Phase-shift
methods. The high lying resonances are found to be not genuine as
pointed out by Kato and Abe, and as such are discarded. This
discomfiture is due to the (i) shallow behavior of the Morse
potential at the outer edge and (ii) inappropriate insertion of
Coulomb tail used in that work. These two deficiencies are now
removed in the present study by finding a modified Morse potential
with steep rise in the outer edge, and joining smoothly to it a
term approximating the Coulomb tail. Calculation of the resonances
for this modified Morse potential using the above two methods of
identification of resonances in our study, yields more than 25
states with angular momenta $0^{+}$ -~$12^{+}$ in the relevant
energy regions. This reaffirms the diatomic-like rotational and
vibrational picture of the nuclear molecular resonances in
$^{12}$C+$^{12}$C system proposed earlier, and shows close
resemblance with the physics of diatomic molecules, a phenomenon
belonging to altogether a different area. It is revealing that the
similarity extends right upto the level of potential which is
Morse type in both the cases. This study also reveals new features
of heavy ion potential.

\newpage

\noindent{\large {\bf  1.~~~Introduction}} \\

Since the discovery of nuclear molecular resonances (NMR) in
$^{12}C+^{12}C$ in 1960 by Bromley et al \cite{Brom60}, and
subsequent observations \cite{Cind81} of such states in many
combinations of projectile and target mainly in low mass systems ,
fully satisfactory understanding of the mechanism of this
phenomenon has not yet been achieved . Numerous attempts both
microscopic and phenomenological have met with limited success
\cite{Satp92}. First microscopic description with the usual optical
potential \cite{Iman68}, could account
for few low lying states in $^{12}C+^{12}C$, which is a very small
part of the rich spectrum consisting about more than 40 states.
Use of similar potential \cite{Sche71} could describe
only some high lying states in the double excitation model. The
band crossing model \cite{Abe79} which is an extension of the
model used in \cite{Iman68}, was successful in describing some high spin resonances only.
Thus microscopically , a holistic description could not be
attained. On the other hand , the resonance data show
characteristic rotation-vibration structure reminiscent of the
spectrum of a diatomic molecule. Iachello \cite{Iach81} postulated
that, like the spectra of diatomic molecule , the phenomena of
resonance is governed by the dipole degrees of freedom
characterized by  the vector $\overline r$, the distance between
the centers of the two colliding  nuclei and consequently the
potential V(r). Since V(r) is not known and difficult to
determine, he bypassed this completely in favour of an algebraic
approach in which the dynamical symmetry group U(4)[ a radius
vector plus three Euler angles] was used to derive the expression
for the energy eigenvalues in terms of the vibrational quantum
number $n$ and angular momentum $L$ as
\begin{equation}
E(n,L)= -D + a(n+1/2) - b(n+1/2)^{2}+cL(L+1),
\end{equation}
where D, a, b and c are parameters to be determined by fitting the
spectra. Similar formula was evolved using repulsive Eckart
potential by Sahu et al \cite{sahu}. Erb and Bromley \cite{Erb81}
fitted 28 resonances to the above expression quite satisfactorily
and obtained the rotational parameter $c$ to be 0.76 MeV which
corresponds to an intrinsic dumbbell configuration consisting of
two touching $^{12}C$ nuclei. This implies an equilibrium
separation of approximately 6.75 fermi for the two $^{12}C$
nuclei. This suggests that if a bonding potential like
non-absorptive  Morse potential between two $^{12}$C nuclei
exists, then it should have longer range than the usual optical
potential.  It conjures a picture of parallelism between the
physics of diatomic molecules and NMR --- two phenomena belonging
to two different areas of physics, namely nuclear physics, and
atomic and molecular physics governed entirely by two different
types of interaction.  That the similarity could extend right upto
the level of potential which is Morse type in both the cases,
would be indeed illuminating.

Since the heavy-ion potential calculated in different models have
similar features like the Morse potential \cite{Mors29} which has
rotation-vibration spectrum, we had represented \cite{Satp86} the
effective bonding potential between two $^{12}C$ ions comprising
both the Coulomb and Nuclear parts by a Morse potential plus a
constant. Then
\begin{equation}
V_{eff}(r)= A+B[\exp{(-2\beta x)}- 2\exp{(-\beta x)}],
\end{equation}
where, $x=(r-R_{0})/R_{0}$ has four parameters $A$, $B$,
$\beta$,$R_{0}$.  The bound-states of this potential are obtained
by solving the corresponding Schr\"odinger equation describing the
relative motion of the two $^{12}$C ions, which yields eigenvalue
spectrum \cite{Flug74} with vibrational quantum number $n$ and
angular momentum $L$ as
\begin{eqnarray}
E(n,L)&=& A-B+ \frac{{\hbar}^{2}}{2\mu R_{0}^{2}}[2\beta
\gamma(n+1/2)-{\beta}^{2}{(n+1/2)}^{2}+L(L+1) \\ \nonumber
&-&3\frac{(\beta-1)}{\beta\gamma}(n+1/2)L(L+1) \\ \nonumber
&-& \frac{9{(\beta-1)}^{2}}{4{\beta}^{4}r^{2}}{(L+1)}^{2}],
\end{eqnarray}
where $\gamma^{2}=2\mu B {R_{0}}^{2}/{\hbar}^{2}$. \\

We fitted the resonance data with the eigenvalue expression Eq.\
(3) of the potential and succeeded in obtaining its parameters as
$A=6.99$ MeV, $B=6.30$ MeV, $\beta=0.957$ and $R_0=6.97$ fm which
is represented as the solid curve EFGH in Fig.\ 1.  We had used
this potential in our study \cite{Satp92,Satp90} to find out the
resonances in $^{12}$C+$^{12}$C system.  We had calculated its
bound and resonance states and had accounted for more than thirty
five states.  This potential does not have the required Coulombic
behaviour in the outer region, which was neglected in the first
instance assuming that it will not have much effect on the bound
and resonance states.  However, subsequently we have added a
Coulomb tail represented by $Z_1Z_2e^2/r$ and recalculated
\cite{Satp92,Satp90} the states with the modified potential and
showed that, for low-lying states, the changes in the result are
small which increase gradually as one moves to high-lying ones.
This trend was very much expected.  The results were quite similar
to those obtained before the insertion of the Coulomb tail.  Thus,
about thirty states were well accounted for and the rotational
vibrational picture of NMR was reaffirmed.  Kato and Abe
\cite{Kato97} have reexamined our results. We would like to state
that, there are as many as 43 states determined by us which lie
inside the pocket of this potential which have been reproduced by
them in there calculation \cite{Kato97}. Couple of
resonances at the top of the well for each partial wave $\it l$
are also reproduced. However , the high lying ones
calculated by us have been found by them to be not genuine
resonances but rather echoes. We do agree that these states are
not genuine. In the present paper we are revisiting this problem,
finding out the deficiencies  and coming out with satisfactory
solution. Our effective potential EFGH (Fig.\ 1), is supposed to
contain both the nuclear and Coulomb interaction. This potential
is expected to be correct in the interior region upto the Coulomb
barrier, beyond which it is not appropriate. When relevant Coulomb
potential is subtracted it goes to zero at G corresponding to
$r=15$ fm which has been identified as the position of the Coulomb
barrier. Beyond this point the two nuclei are separated and the
potential should be Coulombic in nature given by $Z_1Z_2e^2/r$
represented by the dotted curve GI in Fig.\ 1.  We would like to
emphasize that our method of determining the potential by fitting
Eq.\ (3) with the resonance data has yielded the Morse potential
EFG, valid in the interior region upto the Coulomb barrier G.  The
Coulomb tail GI is the well known exterior part of this potential.
The main features of this Morse potential are a long range of 15
fm, depth of 3.45 MeV and repulsive core of
17.45 MeV.\\

We observe that our Morse potential has unphysical behavior being
highly shallow at the outer edge. To be realistic, it must be
steep in that region like the Woods-Saxon potential. The Coulomb
tail of the potential must be smoothly connected without any kink
at the Coulomb barrier which was not the case in the former study
\cite{Satp92,Satp90} where Coulomb tail GI was inserted in
\textit{adhoc} manner (Fig.\ 1), as pointed out in 
\cite{Kato97}. In the present paper, we repair these two
deficiencies by finding a potential which retains all the
important features of our earlier Morse potential like its
long-range, depth, the repulsive soft core, and newly added
features of steep rise near the outer edge, and further a long
range decreasing tail which simulate the Coulomb interaction, in a
natural way as a continuation. We call this potential as modified
Morse potential. To calculate the resonances of this potential in
a consistent and accurate manner, we have used our two methods
specially developed for this purpose \cite{Sahu02}, namely the
Imaginary Test Potential(ITP) and Imaginary Phase Shift(IPS)
methods which have been shown to be very reliable in identifying
the resonances . We show that around 30 resonance states with
$L=0^{+}$- $12^{+}$ are reproduced in the relevant energy regions
observed in experiment, thus reaffirming the diatomic-like
rotation-vibration structure of NMR.

In Sec. 2, we present the construction of the modified Morse
potential from the original Morse potential. Sec.3 describes our
method of calculation of resonances in S-Matrix approach  using
the ITP and IPS methods. In Sec.4, the details of the calculation
and the results are presented. Secs. 5 and 6 respectively give
the discussion on the results, and the conclusions respectively.\\

\noindent{ \large {\bf 2.~Modified Morse Potential}} \\


In this section we would like to modify the effective Morse
potential derived above to make it more realistic. A major
deficiency of this potential is its extreme shallowness in the
region just to the left of the barrier position G. Our experience
on nuclear potential used in different areas of nuclear physics
over the years , has shown that, the potential has to be sharply
rising in this region, quite akin to Woods-Saxon form with a small
diffuseness. Therefore, for the potential to be realistic it
should steeply fall to the left of the point G and smoothly
connect to the Coulomb-like repulsive tail on the right. We have
constructed such a potential which retains all the essential
features like depth, range and repulsive core of our effective
Morse potential EFGI, and additionally incorporate the above two
features. This potential termed as modified Morse potential will
be used in our calculation of resonances which is given by
\begin{eqnarray}
V(r)&=&V_{I}e^{-a_Ir}+V_{0}[\xi_{1}-(\xi_{1}-\xi_{2})\rho_{1}(r)],~~~if ~~r
\leq R_{0} \\ \nonumber
     &=& V_{0}\xi_{2}\rho_2(r),~~~~ if~~ r>R_{0} \\ \nonumber
\end{eqnarray}
where
\begin{equation}
\rho_{n}(r)= [\cosh^{2}\frac{R_{0}-r}{d_{n}}]^{-1}, n=1,2 \nonumber
\end{equation}
and $V_{0}>0 $, has eight parameters $V_{0},\xi_{1},\xi_{2},R_{0},
d_{1},d_{2}$, $V_{I}$, $a_{I}$ and quite flexible to admit varieties of shape.

The modified Morse potential, thus generated, is presented in
Fig.2(b) with the addition of centrifugal term for the angular
momenta $\it l$=$0^{+}-12^{+}$. The parameters are $R_0$=15 fm,
$d_{1}=0.25$, $d_{2}=15$, $V_{0}=$1 MeV, $\xi_{1}=0.45 $,
$\xi_{2}=3.45 $, $V_{I}=$17 MeV and $a_{I}$=0.57 fm$^{-1}$. This
potential for $\it l=0$ has range 15 fermi, depth 3.45 MeV, and
soft core of 17.45 MeV as in original Morse potential. For
comparison, in Fig. 2(a), this potential is shown by a solid curve
and the original effective Morse potential $V_{eff}$  for $\it
l=0$ with the added Coulomb tail shown by a dashed curve.  Thus
the Coulomb tail in the modified Morse potential is only
approximate.

\noindent
{\large {\bf 3. ~Methods of Calculation}} \\

To calculate the resonances, we follow the more versatile S-matrix
approach \cite{Newt66,Tayl72,Alfa75} rather than the phase-shift
method followed earlier in Refs. \cite{Satp92,Satp86,Satp90}. The
partial wave S-matrix in potential scattering is expressed as
\begin{equation}
S_{l}(k)=e^{2i\delta_{l}(k)}=\frac{W[\phi_{l}(k,r),f_{l}(k,r)]e^{i\pi \it l}}
{W[\phi_{l}(k,r),f_{l}(-k,r)]},
\end{equation}
where W[$\phi_{l}(k,r),f_{l}(\pm k,r)$]=$f_{\it l}(\pm k)$ are
known as Jost functions and W is the Wronskian of the regular
$\phi_{\it l}(r)$ and the irregular $f_{\it l}(\pm k,r)$ solution
of the modified Schrodinger equation describing the scattering of
two colliding particles with reduced mass $\mu$,  center of mass
energy E, wave number $k=\sqrt{\frac{2\mu E}{\hbar^{2}}}$ and
phase shift $\delta_{l}(k)$. The resonances are  identified with
the poles of $S_{l}(k)$ arising from the zero $(k_{p})$ of $f_{\it
l}(-k)$ in the fourth quadrant of the complex k-plane such that
$k_{p}=k_{r}-ik_{i}$. The resonance energy is
\begin{equation}
E_{R}=\frac{\hbar^{2}}{2\mu}(k_{r}^{2}-k_{i}^{2}), \nonumber
\end{equation}
and the corresponding width is
\begin{equation}
\Gamma=\frac{\hbar^{2}}{2\mu}4k_{r}k_{i}. \nonumber
\end{equation}
However in practice, the computation of poles is a non-trivial
task, because in the Newton-Raphson type of iteration search
programme generally followed, if the starting trial value of
$k^{2}$ or $k$ is not reasonably close to the resonant pole, the
iteration procedure may not lead to exact pole position. So in
such procedure, one is not sure if one has succeeded in
identifying all the possible resonances present in the system. For
large width resonances, the number of iteration for convergence
may be unusually large. For very narrow resonances, the chances of
skipping the poles in the iteration procedure are quite high.
Similar difficulties may arise in the alternative complex scaling
method \cite{Gyar71,Krup90}. In fact, a recent study by Sahu
\textit{et al} \cite{Sahu04} has shown that nearly half of the
resonances of Morse potential of Satpathy \textit{et al}
calculated by Kato and Abe \cite{Kato97}, using complex scaling
method are not genuine. To circumvent those difficulties, we have
recently developed two methods for unambiguous and authentic
identification of resonances in scattering theory, which we have
employed in the present  study on $^{12}C+^{12}C$ problem. These
are presented in details in Ref.\ \cite{Sahu02}. For easy reading
and completeness we summarize them below.

Our methods are based on the fact that, confirmation of resonances
in collision experiments are ascertained by their correlated
manifestation in different available non-elastic and reaction
channels. We use this concept in the theoretical search of
resonances of a real potential, by adding a small test imaginary
potential $(TIP)$ to the real potential under investigation,
thereby converting the elastic potential scattering problem, to a
two-channel problem involving elastic and absorption channels.
Then in the study of the variation of the partial wave reaction
cross-section with energy, the position of peaks will be
identified as the resonance energy. The method has been termed as
Imaginary Test Potential $(ITP)$ method. This approach eliminates
the difficulties arising from background phase-shift. The resonance
positions can be identified remarkably well in this method, since
in the calculation one can control the height of the absorption
peak in the reaction cross-section by changing the value of the Test
Imaginary Potential. The weak resonances can be located easily
in this method also. 

The second method which is more versatile in the sense that, it
gives both the energy and the width of the resonances, has been
termed as Imaginary Phase Shift (IPS) method. Here we have made
use of the concept of time delay given by
$\tau=\frac{1}{k}\frac{d\delta}{dk}$ (in $k^{-2}$unit) arising due
to the trapping of the incident wave in the resonance
configuration, normally used in the study of resonance phenomena.
By the introduction of TIP, with a strength $W$, the phase shift
becomes complex $\delta=\delta_{r}+i\delta_{i}$ resulting in the
time-delay
\begin{eqnarray}
\tau &=& \frac{1}{k}\frac{d\delta}{dk} \\ \nonumber
  &=&\frac{1}{k}\frac{d\delta_{r}}{dk}+\frac{i}{k}\frac{d\delta_{i}}{dk} \\ \nonumber
&=& \tau_{r}+i\tau_{i}.
\end{eqnarray}
The quantity $\tau_{r}=\frac{1}{k}\frac{d\delta_{r}}{dk}$ has the
usual meaning of time-delay which can be used to estimate the
energy of resonance as it has maximum value at resonance energy.
This method which has been normally used, has inherent uncertainty
due to the requirement of high resolution of energy needed for
differentiation. However, now the imaginary part of the time delay $\tau_{i}$
has been shown to have interesting physical property capable of
giving both the energy $E_{R}$ and the width $\frac{\Gamma}{2}$ of
resonance. We have shown \cite{Sahu02} that resonance energy
$E_{R}$ will be found by the calculation.
\begin{equation}
\frac{d\delta_{i}}{dE}\mid_{E=E_{R}}~=~0, ~~~~~~~~
\frac{d^{2}\delta_{i}}{dE^{2}}\mid_{E=E_{R}}~ <~0, \\
\end{equation}
and the width $\Gamma/2$ by
\begin{equation}
\frac{d\delta_{i}}{dW}\mid_{W=\Gamma/2}~=~0, ~~~~~~~~
\frac{d^{2}\delta_{i}}{dW^{2}}\mid_{W=\Gamma/2}~<~0.
\end{equation}
Thus IPS method is quite powerful to yield well defined values for the energy
of the resonances and their width. In the calculation, the method
is found to be quite easy and reliable, since one has to study only
the variation of the derivative of $\delta_{i}$ as a function of 
${E}$ and/or ${W}$, which is substantially simpler and shorter than
the pole search method or complex scaling method, requiring iterative
search for resonance parameters with the potential for accumulation
of numerical errors.

Although the above two methods are quite satisfactory for large
class of potantials one normally encounters in physical problems,
there are limitations in treating specific cases. The ITP method
is quite fine for identifying the resonances particularly around
the barrier top. For the resonance deep inside the pocket, some
numerical difficulties can arise if the barrier has large height
and width. This is because, in such a case, particularly for lower
energy resonances, the wave function will be highly attenuated in
the interior region, and hence the reaction cross-section may not
show apprecialable peak. In such cases, resonance can be located
by WKB technique used for bound states.

In the IPS method, the calculation of resonance positions is likely
to have same type of problems as stated above since it is
equivalent to ITP method. As for the estimation of width, the method
works quite well for sharp resonances, because in such cases, the effect
of background term is small. The formula (Eq.10-11) is derived using
one level Breit-Wigner expression for resonance amplitude which is only
an approximation. If the width of the resonance is large, its estimation
by this method will not be very satisfactory because of significant
background contribution to partial wave amplitude.  \\

\noindent{\large {\bf  4. ~Calculation of Resonances}} \\

As discussed above, for the identification of resonances we need a
test imaginary potential $V_T$ which we have chosen as
\begin{equation}
V_{T}= -iW_{0}[1-\rho_{1}(r)], ~~~if~~~~ r\leq R_{0}
\end{equation}
with $\rho_{1}$ as defined in Eq.\ (5). We have taken $W_{0}=0.1$
MeV. For different partial waves, we obtain the S-matrix and phase
shifts numerically as function of energy.

Following ITP method, we would like to show first that genuine
resonance states exists at high energy much above the Coulomb
barrier in our potential. For $\it l$=4 and $\it l$=10, we
calculate all the resonances in both the methods. In Fig. 3(a) and
3(b), $\frac{d\delta_{i}}{dE}$ and $\sigma_{R}$ are plotted as
function of E for $\it l$=4  and in Fig. 4(a) and 4(b), the same
quantities for $\it l$=10. It can be seen that the zero crossing
points with negative slopes marked by arrows in Fig. 3(a) and 4(a)
indicating the positions of the resonances, perfectly correlate
with the peaks in the corresponding plot of reaction cross section
$\sigma_{R}$ as a function of E in Fig. 3(b) and 4(b),
respectively. For $\it l$=4, there are 5 observable resonances at
energies 2.46, 3.67, 5.28, 7.29 and 9.6 MeV. The low lying
resonance at 2.46 MeV is not manifested in Fig. 3(a) and3 (b).
However, two of these resonances at 7.29 MeV and 9.6 MeV are found
to be in the energy region above the usual Coulomb barrier which
is about 7 MeV.   It must be stated that, the Coulomb barrier of
our modified Morse potential is 3.45 MeV.  As will be discussed
aposteriori, this potential is generated in the final phase of the
reaction in the exit channel where highly deformed prolate
configuration manifests giving rise to lower barrier.  The
genuineness of these resonances are further ascertained by
plotting the wave function $\mid \psi \mid$ as function of $r$ in
Figs. 5(a) and 5(b) which show the concentration of the wave
function in the interior region of the potential. Similarly for
$\it l$=10, the two high lying states at 7.515 and 10.11 MeV, the
plot of $\mid \psi \mid$ in Figs. 6(a) and 6(b) do confirm the
genuineness of the resonances.

It must be emphasized that, the high lying broad resonances  above
the Coulomb barrier which were absent in our effective Morse
potential, occur here, because of the steep rise of the potential
at the outer edge giving rise to relative accumulation of wave
function in the interior of the potential.

Our search for resonances of the modified Morse potential yielded
37 states with angular momenta $L=0^{+}- 12^{+}$ which are
presented as solid lines in Fig. 7. The circles in the figure
represent the experiment. It is satisfying to find that many of
our calculated resonances lie in the same energy regions as the
experimental spectra. Although quantitative agreement is lacking,
the overall picture emerging from our study can be considered to
be well supported by experiment. This picture described by our
modified Morse potential is entirely due to a real potential. In
the actual collision phenomenon, there must be some associated
absorptive processes however weak they may be. To simulate this
physical feature in our study, we take additionally the weak
imaginary potential $W_0$=0.5 MeV and recalculate the resonances.
The results of such calculation are compared with experiment in
Fig. 8. It can be seen that 9 low lying states with $\it l=0-12$
have become extinct giving rise to much better agreement with
data. In the calculated spectra for all the five angular momenta
$\it l$= $0^{+},2^{+}, 4^{+} ,6^{+}, 8^{+}$, the agreement is
reasonable. Thus we can conclude that as many as 25 states
predicted with our modified Morse potential are relevant and
genuine.  Similar results have been obtained 
in \cite{Sahu03} using a potential with a well followed by a
wall-like thick barrier of about 7.7 fm. However, how such a heavy-ion
potential can be generated using realistic nucleon-nucleon potential,
 has to be seen.\\

\noindent{\large {\bf 5. ~Discussion}} \\

The complexity involved in the description of heavy-ion collision
has been highlighted here.  However, like any other potential
scattering approach, the present approach has attempted to
bye-pass these difficulties, rather overcome it.  The two methods
ITP and IPS developed by us before \cite{Sahu02}, and used here in
the calculation have yielded unambiguous and authentic results in
terms of both the energy and width of the resonances.  The energy
calculated by both these methods exactly match.  Thus, the
uncertainty usually inherent in  the identification of resonances, 
and the possibility of missing some of them, are fully eliminated.
Therefore, we have complete trust on these resonances, and regard
them as the genuine ones of the potential we have used. The
resonances thus generated are qualitative in nature confirming the
experimental trend which is satisfying --- in view of  no adjustment
of parameters to fit the calculated resonances with the data. It is
all the more pleasing that the result is consistent in the sense
that the potential determined from the resonance data using the
eigenvalues of the Schrodinger equation, describes the resonances
calculated in the rigorous S-matrix method. This is indeed parallel to
the scheme followed in the study of nuclear spectra. Therefore,
considering the goodness of the identification and the moderate
quality of the results, we feel a reasonable description of the
resonances of $^{12}$C+$^{12}$C system has been possible.  The
physics implication of this with regard to our potential, and the
resulting mechanism of NMR is discussed below.

\noindent{\bf (a) ~Potential}\\

True understanding of the mechanism of nuclear molecular
resonances can be feasible, only when we have the proper bonding
potential between the two colliding nuclei derived quantum
mechanically based on sound physical principles . Starting with
nucleon-nucleon potential determined  from the experimental
phase-shift analysis, one can fold the density of the colliding
nuclei and obtain the heavy-ion potential. The folding potentials
thus obtained, have been widely used in heavy-ion physics over the
years. The main discomfiture of such potentials is the density
profiles of the colliding nuclei --- which forms an important
element in the calculation --- goes on changing in the course of
collision. Further it is an evolving entity governed by the
reaction dynamics and the bombarding energy. The sudden potential
derived using frozen density approximation, may be too simplistic
to describe a highly complex phenomenon like NMR. The adiabatic
potential pertains to another extreme case which is applicable to
slow process involving low bombarding energy. To overcome this
difficulty, we had followed a pedagogical quantum mechanical
method to determine the potential from the resonance data itself.
Treating the nuclei as a two-body problem with a potential acting
between them, it was shown that a combination of Morse potential
plus a constant is a good representation of the effective
potential in the interior region. The corresponding Schrodinger
equation has analytic solution which yields an eigenvalue spectrum
with rotation-vibration feature. Fitting the resonances with the
eigenvalue expression, the parameters of the bonding potential
could be determined. The problem is quite similar to the two-body
deuteron problem where early informations on n-p potential were
derived by supposing a finite square-well represents the essential
features of the potential namely, the range and depth. It must be
recognized that, because of the amenability of the Schrodinger
Equation for Morse potential to analytic solution, we were
successful in unearthing the underlying potential. Not
withstanding its unrealistic behavior at the outer edge , its main
features like depth, range, and repulsive  soft core have been
obtained. As shown in the previous section, retaining these
features intact, and repairing the outer edge by making it more
steep, and incorporating effective Coulomb-like tail, the
resonances in $^{12}C+^{12}C$ have been described. All the states
have been described as resonances, unlike our previous
description \cite{Satp92}, where the states lying in the potential well were
treated as bound-states, and the states above as resonances. This
potential is called modified Morse as it retained all the
essential features of general Morse potential in the interior
region and only the edge was modified by making it more stiff. It
is quite parallel to our traditional use of harmonic oscillator
potential --- because of its analytic solvability --- in our
nuclear structure physics, with its unrealistic shallow outer
edge. It is usually replaced in special
cases by Woods-Saxon potential which steeply rises in that region.\\

It may be recalled here that the diatomic molecules show prominent
features of rotation and vibration in their spectra which have
been well accounted for since 1929, by the use of Morse potential.
The present study with a modified Morse potential shows that NMR
which are sometimes referred to as nuclear molecules are governed
by similar physics as the diatomic molecules.  While the former
are bound states of two atoms (like O$_2$), the latter are
quasi-bound states of two nuclei, both showing rotation-vibration
features governed by Morse type of potential.  It is indeed
gratifying to see the close resemblance of two different phenomena
pertaining to two different areas of physics, namely the molecular
physics and nuclear physics, governed by electromagnetic and
strong interactions respectively.  The present work shows that the
similarity extends right upto the level of potential which is
Morse type in both the cases.  This is indeed quite revealing and
satisfying.

\noindent{\bf (b) ~Mechanism of resonances }\\

The modified Morse potential determined above, offers a wholesome
understanding of the mechanism of NMR. The potential being
throughout positive for all values of the coordinate $r$, will
support states which are transient in nature having short
life-time like the resonances. Its long-range feature of 15 fm is
compatible with the well-known $3\alpha$ linear chain structure
\cite{Frie71} of the first $0^+$ excited state of $^{12}$C, since
two such nuclei aligned along the symmetry axis can generate such
a potential.  In fact microscopic calculations
\cite{Satp92,Sara90,Faes84} of $^{12}$C--$^{12}$C potential --- in
their highly deformed 3$\alpha$ linear chain configuration --- in
$\alpha$-particle, $\alpha$-cluster and folding models using
Brink-Boeker, Volkov and Ali-Bodmore potentials have amply
justified this feature.  The very fact that the cross sections of
$^{12}$C+$^{12}$C reaction do not conform exclusively to the
picture of two $3\alpha$ chains colliding in the scattering, one
cannot preclude the possibility of the two $^{12}$C nuclei can be
resonating within a long-range Morse-like potential in the exit
channel. The long-range of the potential offer the possibility of
the two nuclei to remain all along in the dinuclear regime and
interact retaining their identity. This gives a natural
explanation for the puzzle to understand how these states which
lie at more than 20 MeV of excitation in the corresponding
compound system $^{24}Mg$ without being washed away in the sea of
high level density. This discomfiture was temporarily overcome in
the past by proposing the hypothesis \cite{Grei71,Arim81} of the
existence of a molecular window with a narrow band of low level
density in the compound nucleus. In the present model, such a
postulate is not necessary as the two nuclei retain their identity
all along, although individually they are likely to get excited to
low-lying prolate-deformed states. In particular in
$^{12}C+^{12}C$ system, the $0^{+}$ state at 7.65 MeV in $^{12}C$
which is a linear chain of three $\alpha$-clusters, has been
supposed by Feshbach \cite{Fesh76} to be excited in the NMR
phenomena. The level density at such excitation is only few levels
per MeV.

The most intriguing feature of the resonance spectrum in a system
is that it is not generally  random in nature. It shows the
characteristics of rotation-vibration similar to the picture seen
in a diatomic molecule. This suggests that the genesis of these
states must belong to a common substratum conducive for the
manifestation of such regularity. The long-range potential is such
a substratum which  can accommodate spatially the colliding pair
of nuclei for  adequate time without merging through fusion
reaction, and thereby favouring the generation of rotation-
vibration spectrum. Thus the natural explanation
\cite{Satp92,Satp86} of the mechanism of NMR proposed earlier, and
is strongly endorsed by the present study, is the following.

In the entrance channel, the two spherical nuclei approach each
other and develop oblate deformation because of Coulomb
interaction. As they come closer either by sub-barrier tunnelling
or overcoming the Coulomb barrier, depending upon their energies,
they interact strongly as a composite system without losing their
identity. In the effort to separate they develop strong prolate
deformation in the exit channel which gives rise to a thick
Coulomb barrier which inhibits separation. Hence they undergo
rotational and vibrational motion like a diatomic molecule and
generate NMR. Finally they separate with the restoration of
original shape in the exit channel. This feature of collision
dynamics has been demonstrated numerically in the surface friction
model of Gross and Satpathy \cite{Gros81,Gros82,Gros82a} before.
This long-range potential does not resemble in any way to the
sudden or adiabatic potential normally used in various studies. It
is visualized to be produced in the final phase of the collision
in the exit channel where the system spends relatively longer time
in a strongly elongated prolate configuration. This may be
considered an effective optical potential in which most of the
other channels have been taken into account except the fusion and
transfer channel.  This potential is the result of the dynamics of
the colliding nuclei and therefore the model has been termed
earlier as Dynamic Potential Model \cite{Satp92}.\\

\noindent{\large {\bf ~~~6. Conclusion }} \\

In summary, we have repaired the two deficiencies of the Morse
potential determined earlier from the resonance data of
$^{12}C+^{12}C$, namely, the shallow behavior of the potential at
the outer edge and the ad-hoc insertion of the Coulomb potential.
We have constructed a modified Morse potential which retains all
the main features of the old Morse potential like the long-range
of 15 fermi, shallow depth and repulsive soft core, and added to
it, the new features of steep outer edge and a smoothly continued
approximate Coulomb tail. The resonances of this potential have
been calculated in the S-matrix approach, using our two newly
developed methods namely, Imaginary Test Potential and Imaginary
Phase Shift methods. Both the energy and width of above 25
resonance states have been determined in our study in a convincing
and authentic way. The calculated states with well defined spin
and parity lie in the same energy region where such states  are
observed  in experiment.

Experimentally more than 40 resonances have been observed, some of
which lie at very high excitation around 20 MeV. The highest state
with L= $12^+$ predicted by our calculation at 15.3 MeV, is
somewhat close to the experiment. The few states above it are
probably produced by some other mechanism and cannot be accounted
for by our modified Morse potential. Nevertheless, since as many
as 25 states have been reasonably well described in our study, it
is fair to conclude that modified Morse potential is appropriate
for a comprehensive explanation of the resonances in
$^{12}C+^{12}C$ system. More importantly it reaffirms the
diatomic- like rotation-vibration picture of NMR as concluded
before.  It is indeed satisfying to note that the present study
shows a close resemblance between the physics of diatomic
molecules and nuclear molecular resonances, though they belong to
different areas governed by different interactions at microscopic
level.  This resemblance becomes all the more striking as it
extends to the level of interaction potential being Morse type
in both the cases.

The mechanism of NMR that has naturally emerged is that, the two
colliding nuclei in their spherical ground state approach one
another in the entrance channel and develop oblate deformation. As
they reach the closest proximity either by overcoming the Coulomb
barrier or sub-barrier tunnelling, a composite system is formed
with individual nucleus retaining its identity. In the exit
channel, they develop strong prolate deformation giving rise to a
wide Coulomb barrier. Being caught behind the barrier they undergo
rotation and vibration in their effort for re-separation and thus
generate NMR. Finally they are separated with the restoration of
original shape. Thus these states are produced in the final phase
of the reaction in the exit channel. This mechanism is distinctly
different from other mechanisms in which these states are produced
in the early phase of the collision in the entrance channel before
undergoing fusion and other absorptive processes. In the present
picture, the entire collision process takes place in the dinuclear
regime. The present potential is not ad-hoc unlike in many
studies. Its main features are determined from the resonance data
itself through a well defined quantum mechanical procedure very
much like n-p potential from the deutron data. The present
rigorous calculation of resonances using the same, and their
agreement with as many as 25 states, resulting in the plausible
explanation of the mechanism shows the study is consistent and
theoretically well founded, which most importantly also reveals
new features of heavy-ion potential.

\noindent{\bf ~~~Acknowledgements} \\
I would like to express gratitude and appreciation to C.S. Shastry
and B. Sahu for extensive discussions, calculation and critical
reading of the manuscript.  The work was done under the financial
support of CSIR, Govt. of India.

\newpage
\begin{figure}
\begin{center}
\leavevmode \hbox{\epsfxsize=6.in \epsfclipon
\centerline{\epsffile{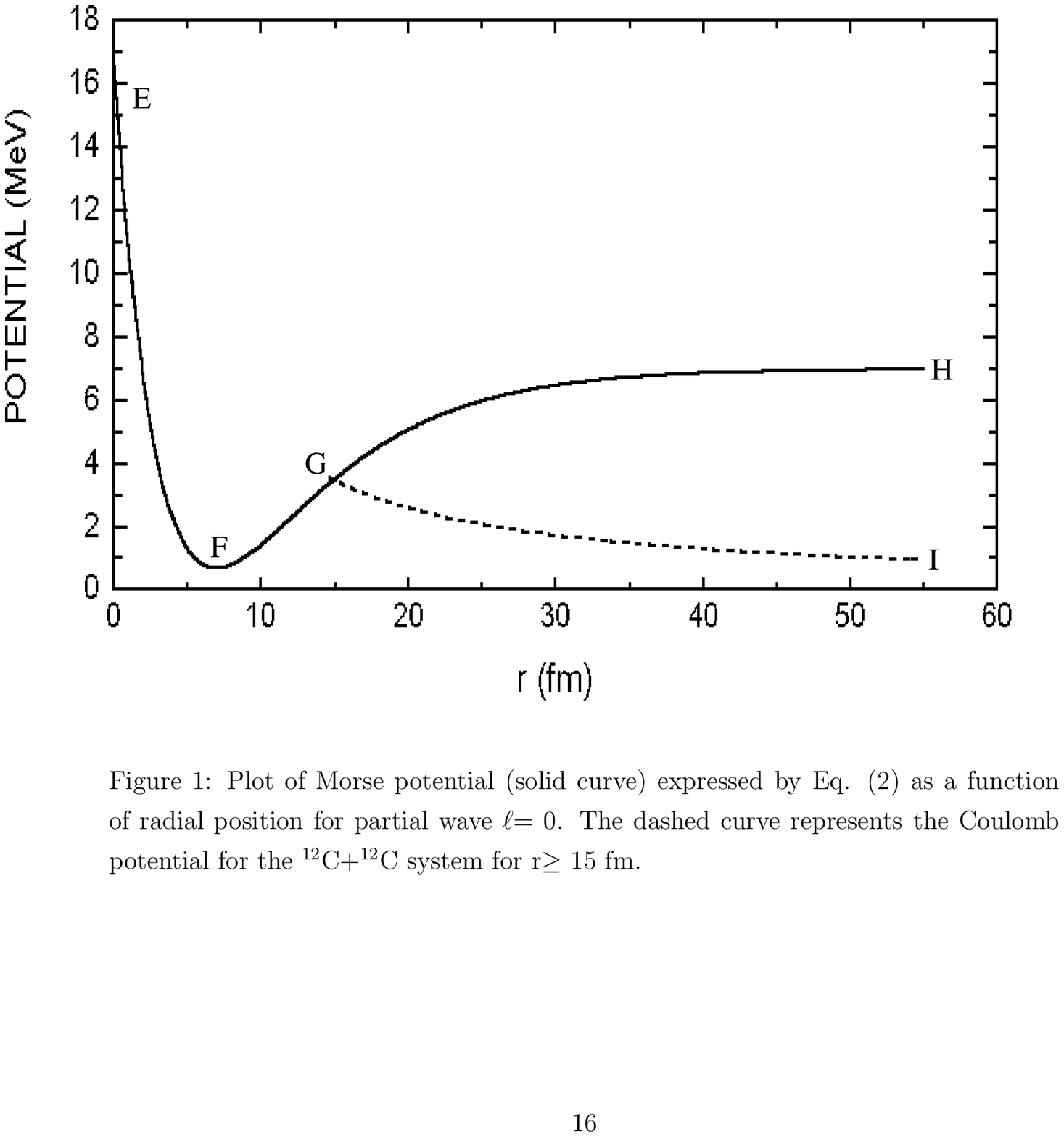}}}
\end{center}
\caption{ Plot of Morse potential (solid curve) expressed by Eq.\
(2) as a function of radial position for partial wave $\ell$= 0.
The dashed curve represents the Coulomb potential for the
$^{12}$C+$^{12}$C system for r$\ge$ 15 fm. }
\end{figure}
\newpage
\begin{figure}
\begin{center}
\leavevmode \hbox{\epsfxsize=5.in
\centerline{\epsffile{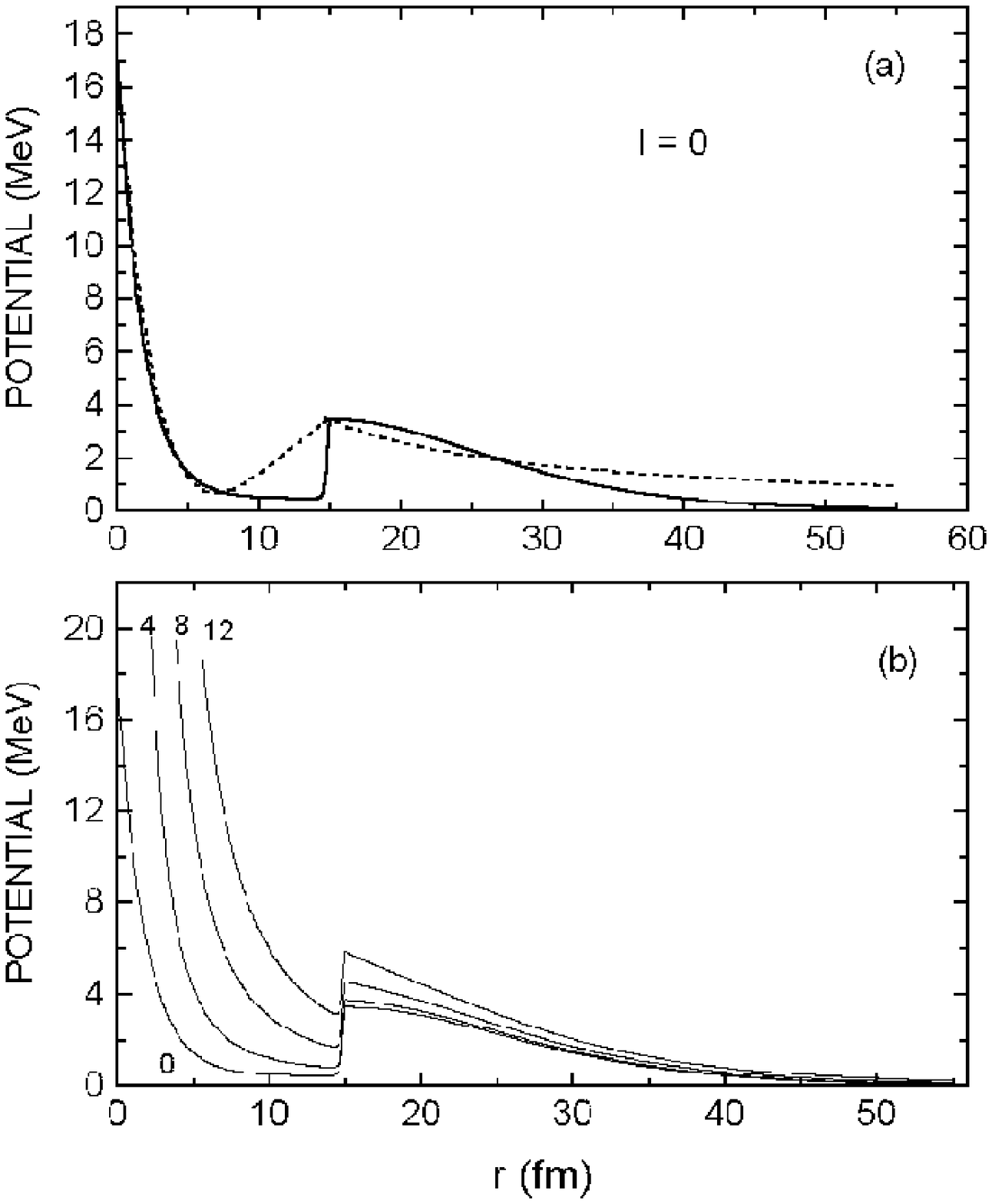}}}
\end{center}
\caption{ Plot of potential as a function of radial position. (a)
Morse potential expressed by (2) with a Coulomb tail beyond r=15
fm is compared with newly constructed potential expressed by Eq.\
(4) for $\ell$=0. (b) The Morse-like potential (4) for different
$\ell$s = 0, 4, 8, 12. }
\end{figure}

\newpage
\begin{figure}
\begin{center}
\leavevmode \hbox{\epsfxsize=6.0in
\centerline{\epsffile{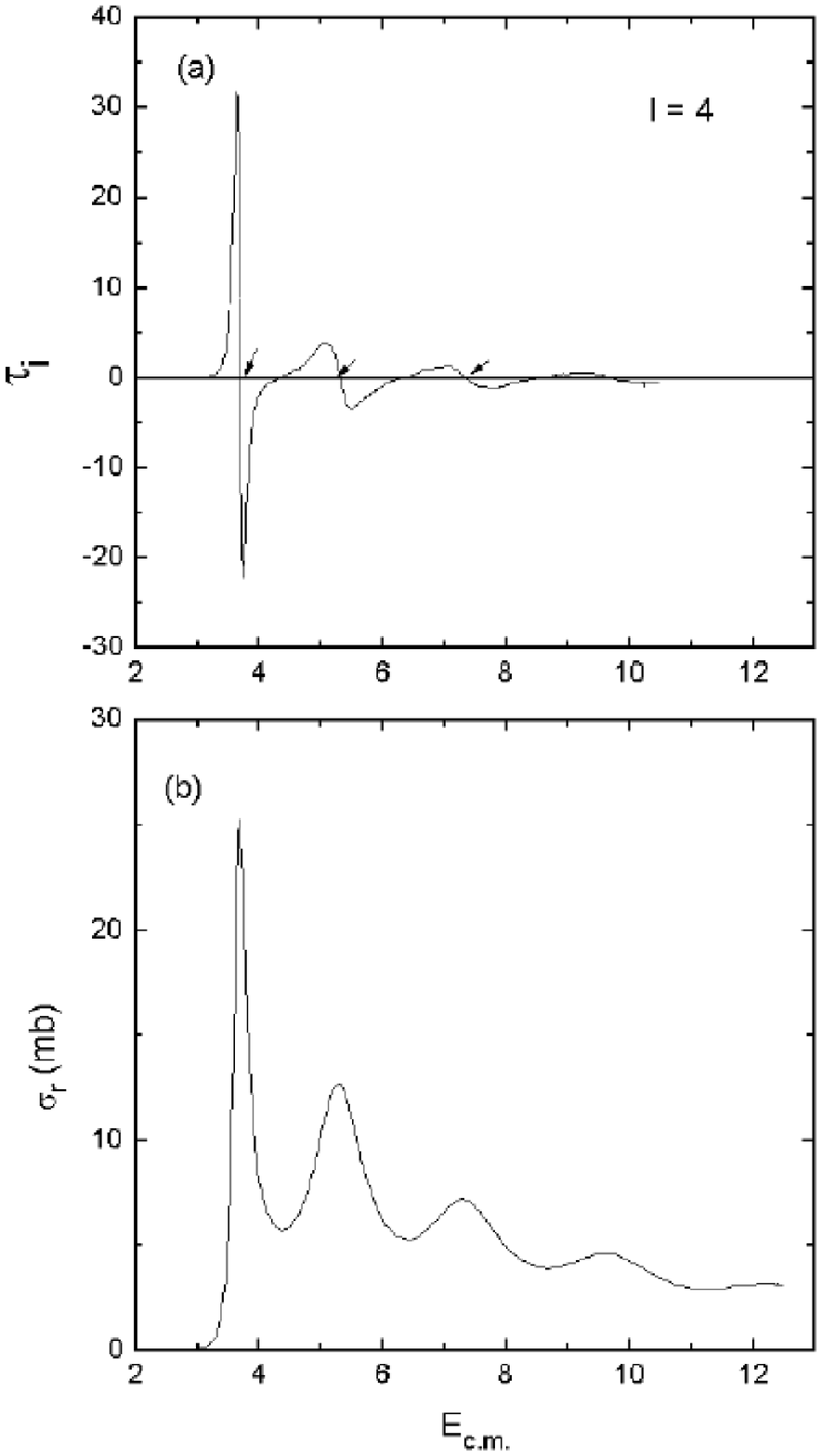}}}
\end{center}
\caption{
(a) Variation of imaginary phase-shift time
$\tau_i$= $\frac{d\delta_i~~}{dE_{cm}}$ as a function of center
of mass energy for $\ell $=4. Arrows indicate zero crossing points with
negative slope indicating  resonance energies.
(b) Variation of reaction cross section $\sigma_r$ as a function of center
of mass energy for $\ell$=4.
}
\end{figure}

\newpage
\begin{figure}
\begin{center}
\leavevmode \hbox{\epsfxsize=6.in
\centerline{\epsffile{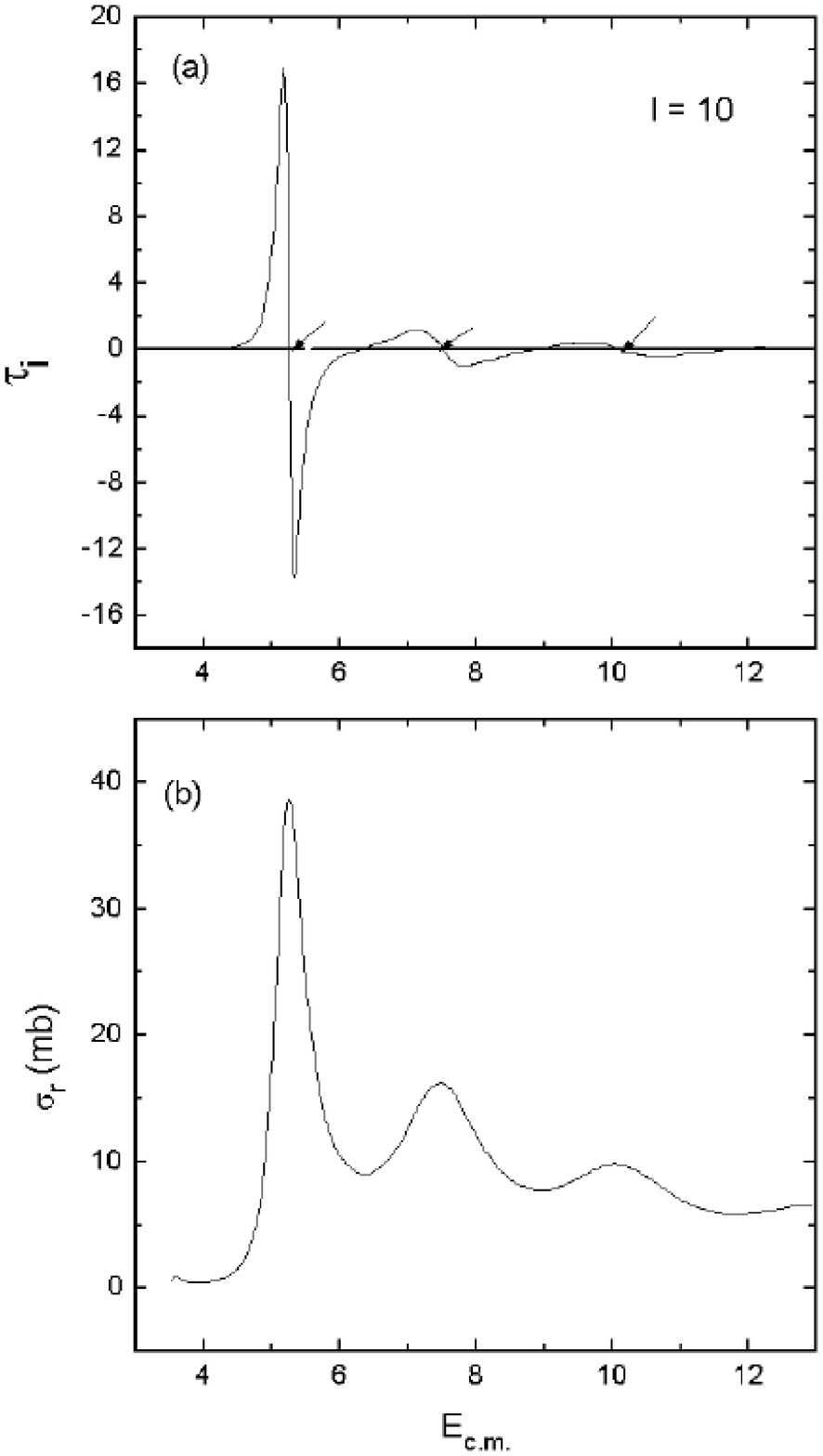}}}
\end{center}
\caption{
(a) Same as Fig. 3(a) for $\ell$=10.
(b) Same as Fig. 3(b) for $\ell$=10.
}
\end{figure}

\newpage
\begin{figure}
\begin{center}
\leavevmode \hbox{\epsfxsize=6.in
\centerline{\epsffile{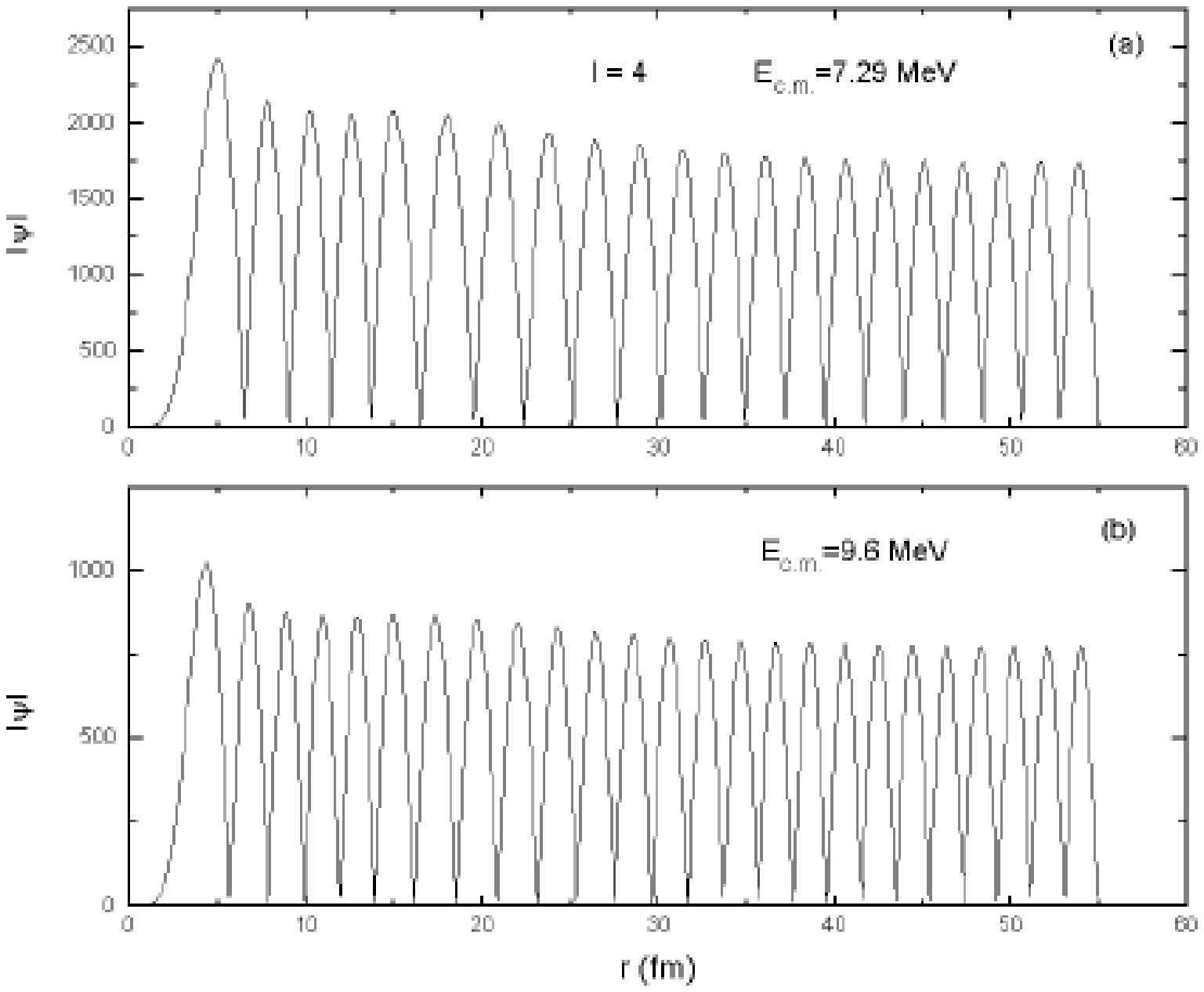}}}
\end{center}
\caption{ Radial variation of modulus of wave function for
$\ell$=4 at resonance energies (a) $E_{c.m.}$=7.29 MeV and (b)
$E_{c.m.}$ =9.6 MeV. }
\end{figure}

\newpage
\begin{figure}
\begin{center}
\leavevmode \hbox{\epsfxsize=6.in
\centerline{\epsffile{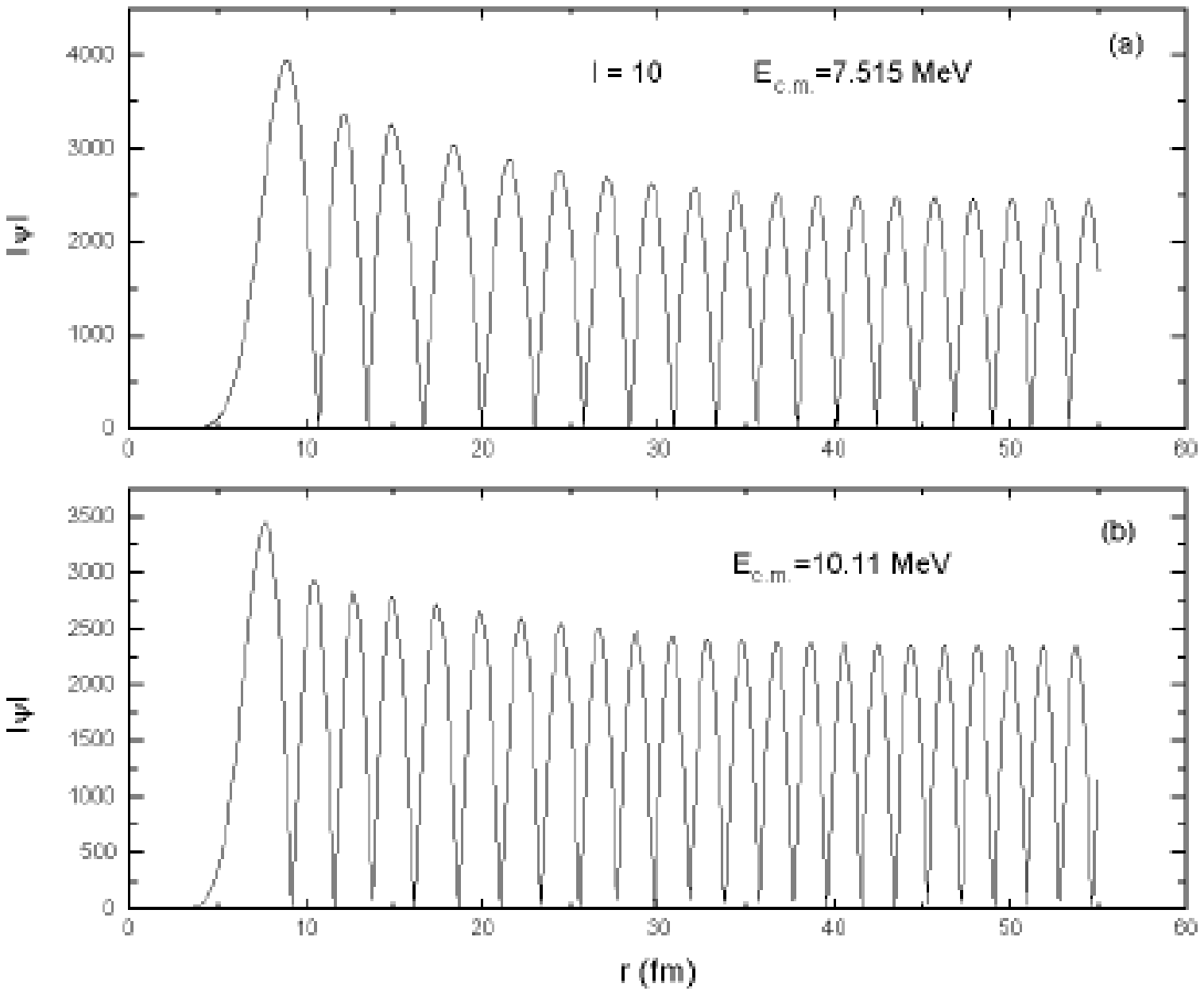}}}
\end{center}
\caption{
Same as Fig. 5 for $\ell$=10 at resonance energies (a) $E_{c.m.}$= 7.515 MeV
and (b) $E_{c.m.}$=10.11 MeV
}
\end{figure}

\newpage
\begin{figure}
\begin{center}
\leavevmode \hbox{\epsfxsize=6.0in
\centerline{\epsffile{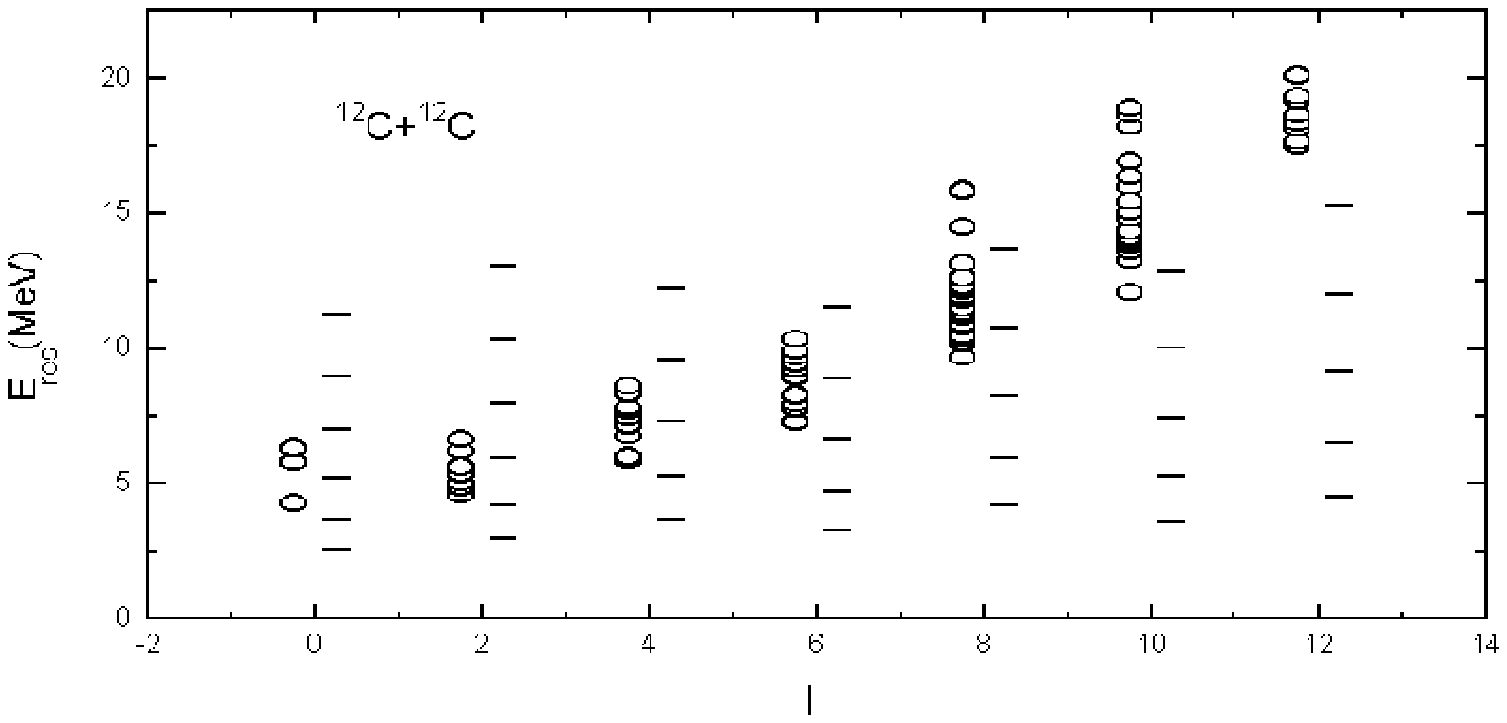}}}
\end{center}
\caption{ Plot of resonance energy for different angular momenta
$\ell$ for $^{12}$C+$^{12}$C system. Present calculated results
are shown by short horizontal lines and the corresponding
experimental results are represented by open circles. } \vskip 3in
\end{figure}

\newpage
\begin{figure}
\begin{center}
\leavevmode \hbox{\epsfxsize=6.0in
\centerline{\epsffile{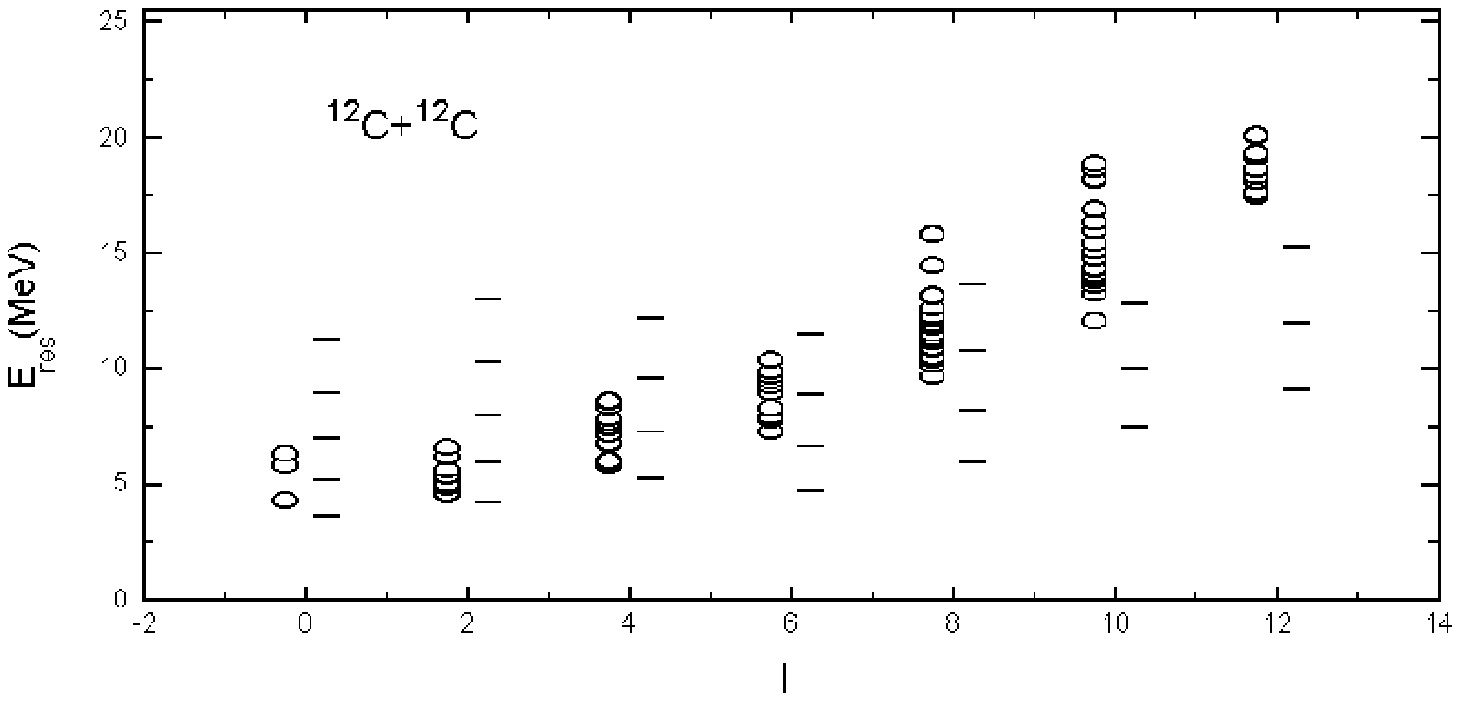}}}
\end{center}
\caption{ Same as Fig.\ 7 but with an additional weak imaginary
potential $W_0=0.5$ MeV. }
\end{figure}

\end{document}